# Rapid suppression of quantum many-body magnetic exciton in doped van der Waals antiferromagnet (Ni,Cd)$PS_3$

Junghyun Kim[1,2#], Woongki Na[3#], Jonghyeon Kim[4#], Pyeongjae Park[1,2], Kaixuan Zhang[1,2], Inho Hwang[1,2], Young-Woo Son[5], Jae Hoon Kim[4,*], Hyeonsik Cheong[3,*], and Je-Geun Park[1,2,*]

[1]Center for Quantum Materials, Seoul National University, Seoul 08826, Republic of Korea

[2]Department of Physics & Astronomy, Seoul National University, Seoul 08826, Republic of Korea

[3]Department of Physics, Sogang University, Seoul 04107, Republic of Korea

[4]Department of Physics, Yonsei University, Seoul 03722, Republic of Korea

[5]School of Computational Sciences, Korea Institute for Advanced Study, Seoul 02455, Republic of Korea

# Authors with equal contribution

* Corresponding author: super@yonsei.ac.kr; hcheong@sogang.ac.kr; jgpark10@snu.ac.kr

**Abstract**

The unique discovery of magnetic exciton in van der Waals antiferromagnet $NiPS_3$ arises between two quantum many-body states of a Zhang-Rice singlet excited state and a Zhang-Rice triplet ground state. Simultaneously, the spectral width of photoluminescence originating from this exciton is exceedingly narrow as 0.4 meV. These extraordinary properties, including the extreme coherence of the magnetic exciton in $NiPS_3$, beg many questions. We studied doping effects using $Ni_{1-x}Cd_xPS_3$ using two experimental techniques and theoretical studies. Our experimental results show that the magnetic exciton is drastically suppressed upon a few % Cd doping. All these happen while the width of the exciton only gradually increases, and the antiferromagnetic ground state is robust. These results highlight the lattice uniformity's hidden importance as a prerequisite for coherent magnetic exciton. Finally, an exciting scenario emerges: the broken charge transfer forbids the otherwise uniform formation of the coherent magnetic exciton in (Ni,Cd)$PS_3$.

Exciton is the bosonic excitation of electron-hole pairs first proposed in the 1930s by Frenkel.[1] It has since been studied in various physics contexts, such as Bose-Einstein condensation, superconductivity and many-body physics, to name only a few. In a simple nonmagnetic band insulator, such as the two-dimensional (2D) transition metal dichalcogenides (TMDCs) without a magnetic ordering, the bound electron-hole pair features considerable binding energy and a long exciton diffusion length for realizing the exciton-based devices.[2, 3] By comparison, it is relatively rare in magnetic insulators, which otherwise would offer an interesting case of exciton coupled with electron correlation in the name of Hubbard U.

The recent discovery of magnetic van der Waals materials offers a new direction in studying magnetic exciton as they share very similar crystal features as TMDCs.[4, 5] Indeed, a new type of magnetic exciton has been recently found in $NiPS_3$, van der Waals antiferromagnet of XXZ-type.[6, 7, 8] The magnetic exciton peak in $NiPS_3$ was observed by photoluminescence (PL), optical absorption and resonant inelastic x-ray scattering (RIXS).[6] As a member of the transition metal phosphorous trichalcogenides ($TMPX_3$, TM: 3d transition metals and X: chalcogens) family, $NiPS_3$ is a self-doped charge-transfer insulator[9], hosting both rich magnetic and excitonic properties.[5, 10] Most interestingly, the exciton originates from the transition between two quantum many-body states, i.e. Zhang-Rice singlet (ZRS) and triplet (ZRT) states.[11] It is also important to note that this exciton becomes bright only below the Neel temperature of ~ 154 K. This temperature dependence supports that magnetism plays an essential role in making the exciton optically bright as the ZRS and ZRT states should be present regardless of whether the ground state is paramagnetic or antiferromagnetic.

The narrow linewidth of the exciton peak is another intriguing characteristic: it has the full-width-half-maximum (FWHM) of ~ 0.4 meV,[6] the narrowest ever reported for electronic transition, even narrower than an impurity state. Notably, the coherence, i.e. the simultaneous luminescence from the magnetically ordered $NiS_6$ clusters, develops only below about 50 K when the antiferromagnetic state becomes a stable ground state. It is probably important to note that this width is smaller than the thermal fluctuation of $k_BT$ even when it is cooled below 50 K. There is strong evidence that some form of coherence develops at low temperatures in $NiPS_3$, although the nature of this coherence is enshrined in mystery at the moment. Therefore, one question urgently needs to be addressed on how the coherence relates to the stable magnetic ground state of XXZ type. So the puzzle is whether the coherent exciton requires coherence in the magnon spectrum or/and lattice uniformity. An answer to this question will put some constraints on the correct theoretical model of the magnetic exciton in $NiPS_3$.

In the present work, we examined the doping dependence of the magnetic exciton by replacing Ni atoms with nonmagnetic Cd dopants in the honeycomb lattice. $CdPS_3$ is a nonmagnetic insulator which shares the same crystal structure as $NiPS_3$. We systematically investigated the magnetic properties and the magnetic exciton in $Ni_{1-x}Cd_xPS_3$ while ensuring chemical homogeneity across the sample on the length scale of a few tens of microns. Our optical studies, both PL and optical absorption, reveal a striking effect: doping, at a level of a few %, is sufficient to destroy much of the magnetic exciton. All this happens while there is only a gradual increase in the linewidth. In contrast, the magnetic ground state remains as robust as in the undoped case, except for a slight drop in the transition temperature. These results, supported by our theoretical calculation of magnon dispersion, answer the above question: lattice uniformity is another critical factor to the coherent exciton state residing in the tuned antiferromagnetic ground state.

Figure 1b shows the dopant concentration of the synthesized $Ni_{1-x}Cd_xPS_3$ crystals, synthesized by solid-state reaction and following the chemical vapor transport method, as confirmed by the Electron Dispersive X-ray spectroscopy (EDX) (See Supporting Information for details). The error bar is about ± 0.05 % ~ 1.25 %, indicating the high homogeneity across our samples. The micro-Raman mapping in Figure 1c further verifies the homogeneity of the concentration in both pure and doped crystals. The micro-Raman mapping image was drawn for the P2 peak (See Supporting Information), the satellite Raman peak used for magnetic studies.[10] We measured the powder X-ray diffractometer (XRD) pattern for all the crystals, which all show sharp (00L) Bragg peaks consistent with the space group of C2/m (Figure 1d). The value of FWHM is as small as ~ 0.14° and barely changes across doping up to 10 % (Figure 1e). Thus, the doped crystals have as good crystallinity as a pure $NiPS_3$ crystal. It is to be noted that upon increasing the dopant concentration up to 10 %, the XRD Bragg peaks only shift gradually but with no extra peak. We made the le-Bail fitting of the XRD data to extract the peak position and the corresponding lattice constant precisely (Figure 1e). The lattice constant c increases linearly following the Vegard's law, changing by ~ 0.6 % from x = 0 to x = 0.1. Thus, the gradual increment of the lattice constant indicates the expansion of the $NiS_6$ cluster with no structural transition.

We also measured the temperature-dependent magnetic susceptibility of all the pure and doped crystals. Since the magnetic easy axis is along the in-plane direction, we measured the in-plane magnetic susceptibility with the magnetic field of 1 T applied parallel to the honeycomb plane. The antiferromagnetic phase transition was observed in all the doped systems and a pure crystal. The Neel temperature was estimated from a peak in the temperature derivatives of the magnetic susceptibility. The paramagnetic-to-antiferromagnetic transition temperature (in Figure 2b) almost linearly decreases upon increasing dopant concentration, changing by about 24 % from x = 0 to 0.1. The doping-induced drop in the transition temperature agrees with the percolation theory.[12] Additionally, the Curie-tail or susceptibility upturn at low temperatures gradually increases due to the disordered spins near the nonmagnetic Cd atoms embedded in the background of original collinear-ordered states.

As aforementioned, the extremely narrow and high-intensity peak is key evidence for the coherent phenomena from the simultaneous dipole transition of the local $NiS_6$ clusters. We performed temperature-dependent PL and optical absorption experiments to study the coherent phenomena of the magnetic exciton in all our samples. The PL was conducted from room temperature to the base temperature of 8 K, and the optical absorption was conducted down to 4 K. We also examined several crystals with the same composition carefully to ensure the reproducibility of the experimental results for all the pure and doped crystals. The absorption coefficient was normalized by the sample thickness. As shown in Figure 3a-b, both PL and absorption measurements show the magnetic exciton peak and its sidebands appearing below the Neel temperature. All the spectra were fitted with a Lorentzian function, representing the exciton peak, sidebands, and background (Supporting Information). As the dopant concentration increases, the central exciton peak and the sidebands shift in the opposite energy directions and become undistinguishable at high concentrations at ~ 8 %. The peak energy shift behaved almost identically for both PL and absorption experiments, proving the consistency of the data. However, we did not observe any new PL or absorption peak induced by the nonmagnetic doping.

What is most striking about all the data is how sensitive the magnetic exciton is to doping. Figure 3d shows that the FWHM of the exciton peak broadens almost linearly, indicating the decoherence of magnetic exciton, i.e. loss of the high narrowness. More importantly, the drastic peak integration drop indicates the effective suppression of the original ZRS-ZRT states (Figure 3e). The primary peak amplitude loses most of the initial intensity within a low concentration of ~3 % and reaches nearly zero at 10 % of the Cd doping. Furthermore, it shows that the density ratio of the deformed ZRT-ZRS increases. Additionally, the broadening effect in the linewidth from the less coherent peak in the doped system means the instability of the original cluster. In contrast, the linewidth gradually increases with doping, reaching 5 meV at 8 K for 10% doping.

The color plots of the integrated peak in Figure 3f-g demonstrate the significant integrated peak suppression as a function of the dopant concentration for both the PL and absorption experiments. Although the magnetic exciton peak appeared right below the Neel temperature, we can see the particular bright region, which we can call a coherent region, with the narrowest linewidth as a function of temperature. Generally, the magnetic exciton is correlated to the magnetic states, and the Neel temperature reflects the strength of the magnetic states or the magnon in the doped system. However, in our experiments, the integrated peak in the coherent region showed a drastic change as a function of Cd concentration, while the Neel temperature varies relatively slowly. With this unexpected result, we suspect there should be other factors working as a primary source to the exciton's suppression in addition to the magnon's contribution.

Before jumping to the main account of exciton's suppression, we would like to discuss the magnon's contribution for a better understanding. As shown in Figure 4, the magnon spectra of $Ni_{1-x}Cd_xPS_3$ along the ($H$, 0, 0) direction were calculated using linear spin wave theory, with the spin Hamiltonian based on an inelastic neutron scattering study[13] (see Supporting Information for details). For a fair comparison, the magnon's linewidth is set to be the same as the exciton's for pure $NiPS_3$, and we check the evolution of the magnon's linewidth for doped systems. As the dopant concentration increases, the flat band comes from the localized dopant in real space (Figure 4a-f).

The magnon peaks from the Q-cut at each $\Gamma_0$ and Z are fitted with the Gaussian function to check the linewidth's evolution (Figure 4g-h). The fitted FWHM of the magnon peak for $\Gamma_0$ broadens linearly at a low doping regime and shows no significant change at a high doping regime with increasing dopant concentration. In contrast, the Z peak broadens linearly through all ranges (Figure 4i). As a passing remark, our magnon's suppression is much more prominent than other collinear magnetically ordered systems.[14] Interestingly, the exciton for both PL/absorption and the magnon at Z exhibit similar linewidth broadening behavior, especially at the low doping regime. It implies that the magnon may contribute to the decoherence of the exciton due to the exciton-magnon coupling, as we disturb the magnetic order by nonmagnetic doping. However, our simulated magnon's linewidth differs from the experimental exciton's width, especially at a high doping regime. It is probably because our simulation assumes all the Ni spins to be collinear for all the doped systems as pure $NiPS_3$. In contrast, the doping-induced disordered spins (evidenced by the susceptibility upturn at low temperatures for higher dopings in Figure 2a) may slightly change the original collinear ordering in the high-doping system.

Note that, upon Cd doping, neighboring Ni spins become perturbed from the initial collinear structure, driven by disorder in the spin alignment due to the disrupted exchange interaction. Consequently, the magnon's broadening directly indicates changes in the perturbed magnetic state similar to the Neel temperature. Then, it would lead to changes in magnon, which, in principle, can disrupt the formation of the ZR excitons. In a simple magnon-dominated picture, both magnon and exciton broadening are crucially influenced by the perturbed magnetic state. It's also worth noting that the Neel temperature, similar to the magnon, is an additional direct indicator of disrupted spin states.

Since the Neel temperature decreases only slightly with doping (Figure 3f-g), the magnetic ground state should be crucial in creating the magnetic exciton in the doped systems. As the Neel temperature results indicate, this robust magnetic state implies the stable magnon in the presence of Cd doping. Considering the behavior of the magnon and exciton broadening at the low doping regime, this suggests the magnon scenario alone cannot fully explain the rapid initial suppression of the magnetic exciton. Therefore, the magnon's contribution to exciton's suppression should be limited. Thus, we explain the rapid suppression of the exciton by going beyond the robust magnetic ground state. This leads us to explore other possible accounts for exciton's strong suppression. For this, we emphasize that the magnetic exciton originates from the transition between the multiplets of Zhang-Rice states.

Let us now think about the theoretical implications of our experimental findings. First, we would like to point out that we can safely assume that the antiferromagnetic ground state is robust and stable at the base temperature. Our data show only a tiny drop in the transition temperature and the calculated magnon dispersion, particularly its linewidth as a function of doping. Therefore, the loss of the coherent exciton by a few % doping cannot be explained by imagined or otherwise disruption in the magnon spectrum. Second, the absorption data is less sensitive to the defect/dopant than PL, making it a more reliable measure of the optical oscillator strength. Although we interpret the absorption data mainly to explain the doped system, both data are complementary.

One should also note that the exciton in $NiPS_3$ is a Frenkel type, which makes it quite sensitive to local environments. As discussed, the Frenkel-type exciton considered in this work should have spin components to be optically active. In other words, the Zhang-Rice multiplets can be excitonic states only if the d-state on metal ion and resonating p-states on ligands should have the opposite (same) spin orientation for the ZR singlet (triplet) state. If not, the state should not have proper optically active channels for optical probe experiments. Thus, our exciton involves a local magnetic excitation such that the Cd dopant's quenched local magnetic moment deteriorates its creation around the dopant.

Furthermore, with the eradication of exciton right at the Cd site, variation in interlayer coupling near the dopant will also change the potential energy landscape for the exciton created in the adjacent layers. Because of these aspects, it is natural to assume that excitonic bound energy will be distributed further than without doping. (See Supporting Information) This is also consistent with our experimental observation of the expansion of interlayer distance shown in Figure 1e. So, if we consider the relatively weak perturbation to charge sectors by Cd doping, the FWHM of excitonic peaks would be broadened linearly as a function of the doping ratio, $\sigma(x)/\sigma_0 \approx 1 + \alpha x$: where $x$ is a doping ratio from 0 to ~0.1, $\sigma(x)$ a doping-dependent FWHM, $\alpha$ a proportional constant, and $\sigma_0$ an FWHM without doping. Owing to the broadening, we can similarly expect that the absorption peak integration for the excitons whose bound energies have a doping-dependent distribution will be reduced according to $1/\sigma(x)$. So, per a given density of exciton, its doping-dependent integrated peak intensity of $I(x)$ can be written as $I(x)/I_0 \approx 1 - \alpha x$, where $I_0$ is integrated peak intensity without doping. Unlike a typical Frenkel exciton, as we already discussed, the density of bright excitons should decrease with the doping of nonmagnetic Cd atoms because of its orbital-spin entangled nature.[6] Therefore, the doping-dependent exciton density may be written as $n(x)/n_0 = f_R(x)$ where $n(x)$ and $n_0$ are a density of excitons with and without doping, respectively. Here we introduce a doping-dependent function of $f_R(x)$ where $f_R(0) = 1$ and $df_R(x)/dx < 0$. Then a total integrated peak intensity of bright excitons of $I_T(x)$ with doping can be written as $I_T(x)/I_{T0} \approx f_R(x)(1 - \alpha x)$ where $I_{T0}$ is a total integrated peak intensity without doping.

If we assume further that the quenched ratio of magnetic exciton linearly increases as doping increases, we can approximately write $f_R(x) \sim 1 - \alpha x$. Then, the total integrated peak intensity of the absorption coefficient can be expressed as $I_T(x)/I_{T0} \sim 1 - 2\alpha x$ such that it decreases twice faster than the increasing ratio of FWHM. Therefore, the absorption peak integration should be completely quenched when the FWHM is broadened by 50%. As shown in Figure 3d-e, the case above matches the sample with $x \sim 0.03$. Even though the doping ratio for quenching excitons agrees with our phenomenological considerations, as seen in Figure 3e, the decreasing absorption does not show a linear dependence with $x$. Instead, they get quenched much faster, implying that our bright excitons are fragile. In other words, once Cd dopants are introduced, the exciton around the dopant and away from it seems to quench its brightness sensitively. For the moment, the origin of the apparent fragile excitons cannot be understood well. Considering negligible changes in the linewidth of the exciton upon three orders of magnitude increment of the laser power (See Supporting Information), laser-induced lattice heating or a typical neutral exciton-exciton interactions[15,16] seem irrelevant to our case. We note that, in $NiPS_3$, upon laser illumination, bright spin-orbital entangled excitons are generated with effective local dipole moments, resulting in bright excitonic signals.[6] Our phenomenological consideration on switching-off exciton formation near Cd dopants clearly shows the twice faster quenching of absorption peak than the decrement of an exciton lifetime, agreeing with our observation. This highlights the pivotal role of spin-orbit entanglement in exciton formation. Therefore, we conjecture that the Cd dopant may prohibit magnetic exciton itself, together with a possible role of lattice modulation in breaking coherence of exciton peak because of lattice deformation induced by Cd dopants, leading to rapid quenching of peak integration of the absorption.

Before finishing, we would like to make more general comments. The highly narrow exciton found in $NiPS_3$ has attracted considerable attention, and quantum many-body physics lies at the heart of the problem, making it not easy to address the key questions. What we have done here is to examine how coherence responds to the introduction of either magnetic and/or chemical disorder. First, our experimental results prove that the exciton in $NiPS_3$ is extremely sensitive to nonmagnetic impurities. This experimental observation indicates that the excitons are strongly correlated. The question is, what provides such a strong correlation? Our study on the quenching of this exciton due to Cd doping reveals that the coherence originates from the excitonic state itself. To be more precise, Ni ions in $NiPS_3$ are subject to the near zero or small positive charge transfer energy, leading to the superposition of different Hilbert spaces, namely $d^8$ and $d^9\underline{L}$ states. Coherence is, therefore, observed when the d-state on the metal ion and the resonating p-states on the ligands exhibit the appropriate spin orientation for the Zhang-Rice states. However, lattice inhomogeneity induced by nonmagnetic dopants renders the resulting state optically inactive. This is because the dopant can no longer facilitate proper charge transfer to the shared ligand states in conjunction with adjacent $NiS_6$ clusters.

In summary, we investigated the doping dependence of the magnetic exciton on $Ni_{1-x}Cd_xPS_3$ systems using both PL and optical absorption techniques. All our examples were characterized extensively using several bulk measurements and microanalysis, ensuring chemical homogeneity on a tens-micron scale. The main result is that the highly narrow exciton peak seen in pure $NiPS_3$ loses its coherence rapidly with doping: it is almost destroyed by a few % of Cd doping, leaving a broad sign for higher doping samples. All this happens while the magnetic ground state is as robust and stable. This rapid destruction seems related to our multi-facet excitonic excitations possessing electric and magnetic degrees of freedom.

**Associated content**

Supporting Information: Methods, full spectrum of Raman spectroscopy, derivatives of the magnetic susceptibility, temperature dependence of PL and absorption spectra, fitted exciton peak of PL and absorption spectra, laser power dependence of PL, comparison of Neel temperature of multi-probes, color plot of temperature and dopant concentration dependence of FWHM, simulation of ground-state spin configuration and schematics of magnetic exciton in the doped system.

**Acknowledgment**

We acknowledge many helpful and exciting discussions with Youjin Lee and Suhan Son during the project. The works at Seoul National University were supported by the Leading Researcher Program of the National Research Foundation of Korea (Grant No. 2020R1A3B2079375). Works at Sogang University and Yonsei University were supported by the National Research Foundation (NRF) grant funded by the Korean government (MSIT) (2019R1A2C3006189; 2017R1A5A1014862, SRC program: vdWMRC center; 2021R1A2C3004989). Jonghyeon Kim was supported by the Basic Science Research Program through the National Research Foundation of Korea (NRF) funded by the Ministry of Education (2022R1I1A1A01063982). Y.-W. S. was partially supported by the KIAS individual Grant No. CG031509.

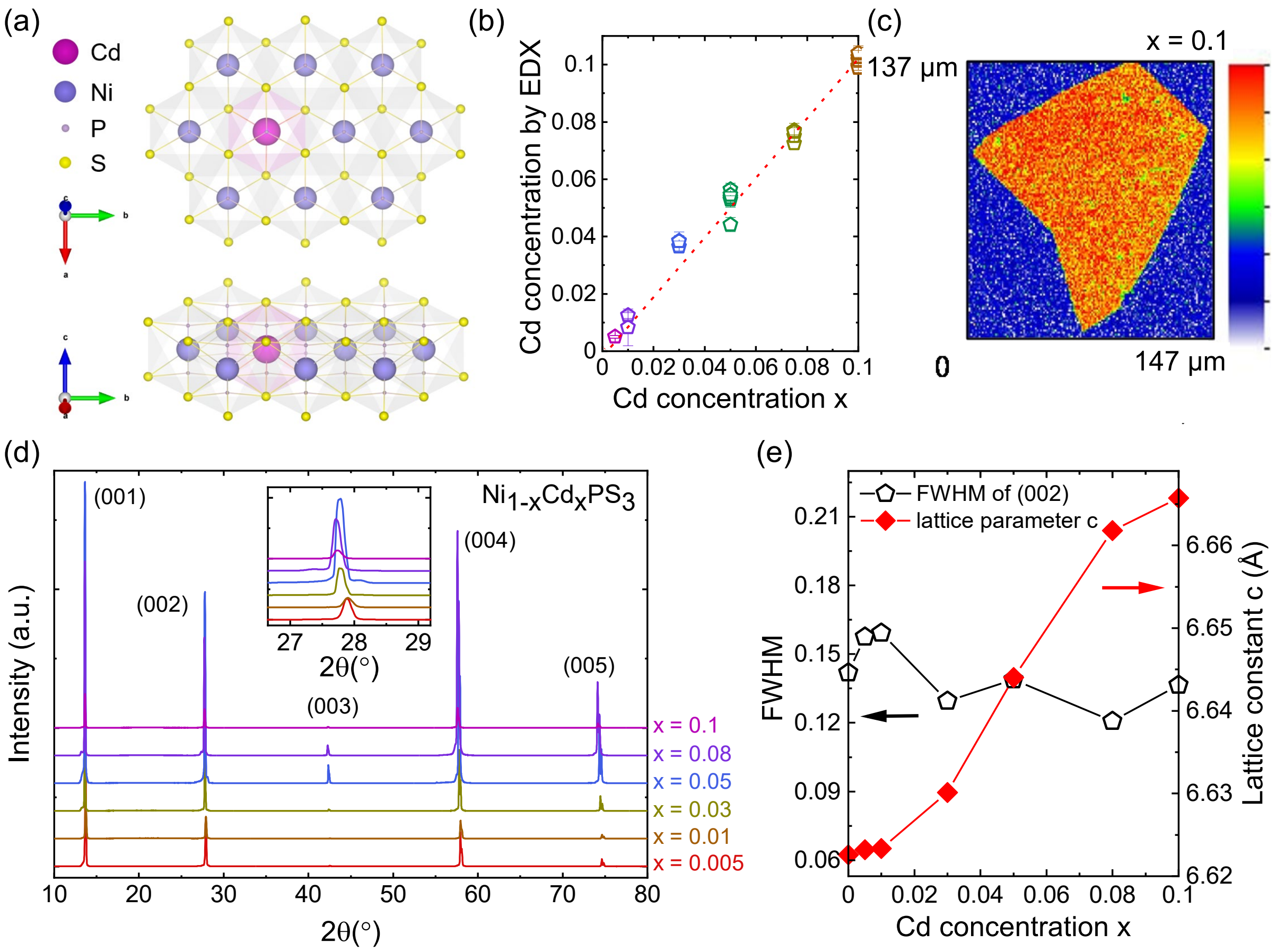

**Figure 1.** (a) Crystal structure of $Ni_{1-x}Cd_xPS_3$. The yellow dots indicate the sulfur atoms. (b) EDX data of the doped systems corresponding to the starting synthesis stoichiometric ratio x for Cd. (c) Image of the micro-Raman mapping of P2 peak (Supporting Information) of the nanoflake for x = 0.1 exfoliated on the $SiO_2$ substrate. (d) XRD data for $Ni_{1-x}Cd_xPS_3$ systems. (e) Lattice constant c of the doped systems from the le Bail fitting of the Bragg peak (002) and the FWHM of the (002) peak.

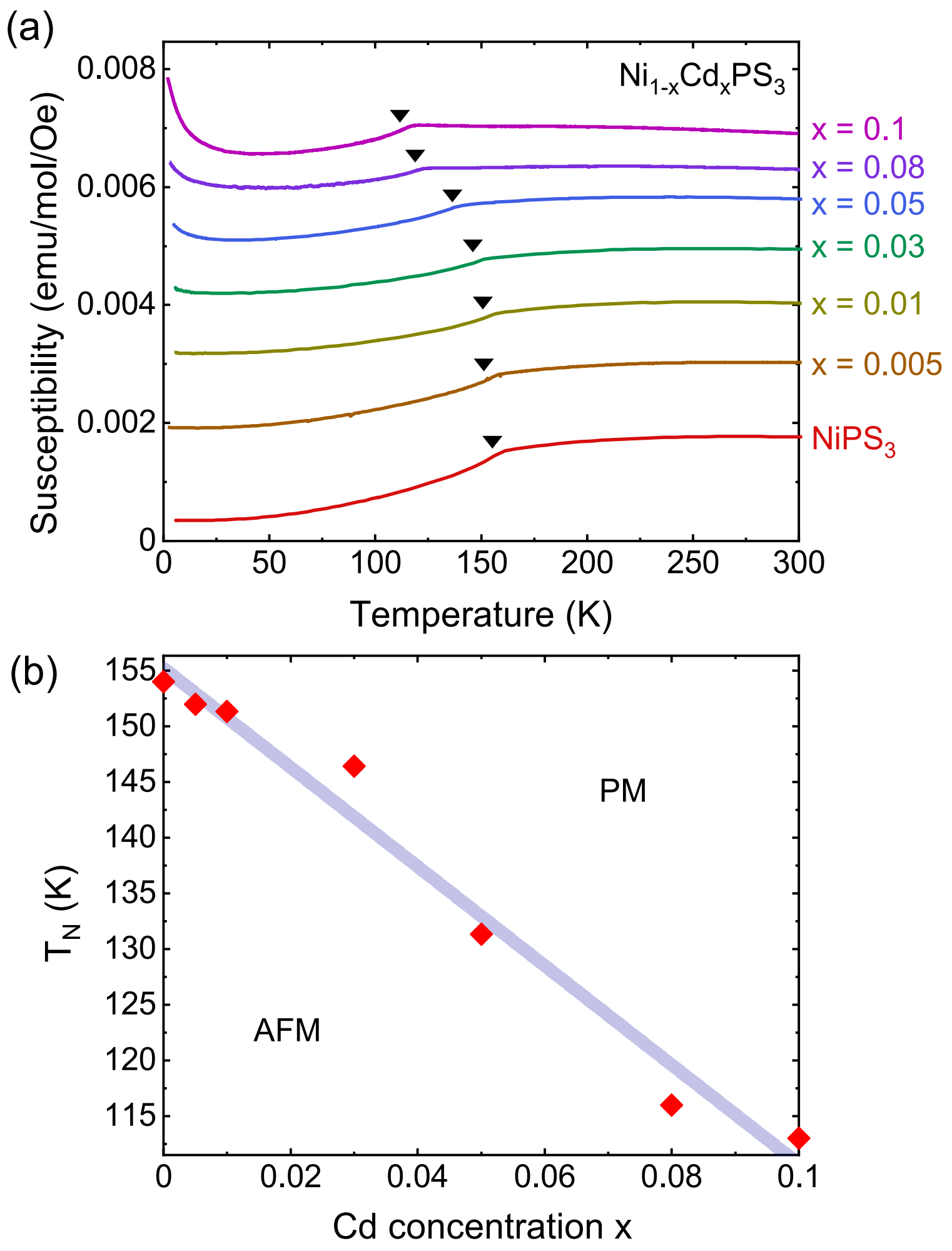

**Figure 2.** (a) Magnetic susceptibility data were taken at $H = 1$ T along the in-plane direction for $Ni_{1-x}Cd_xPS_3$ systems. (b) Neel temperature data was obtained from the magnetic susceptibility of the doped systems, showing the phase transition from the paramagnetic (PM) to the antiferromagnetic (AFM) phase.

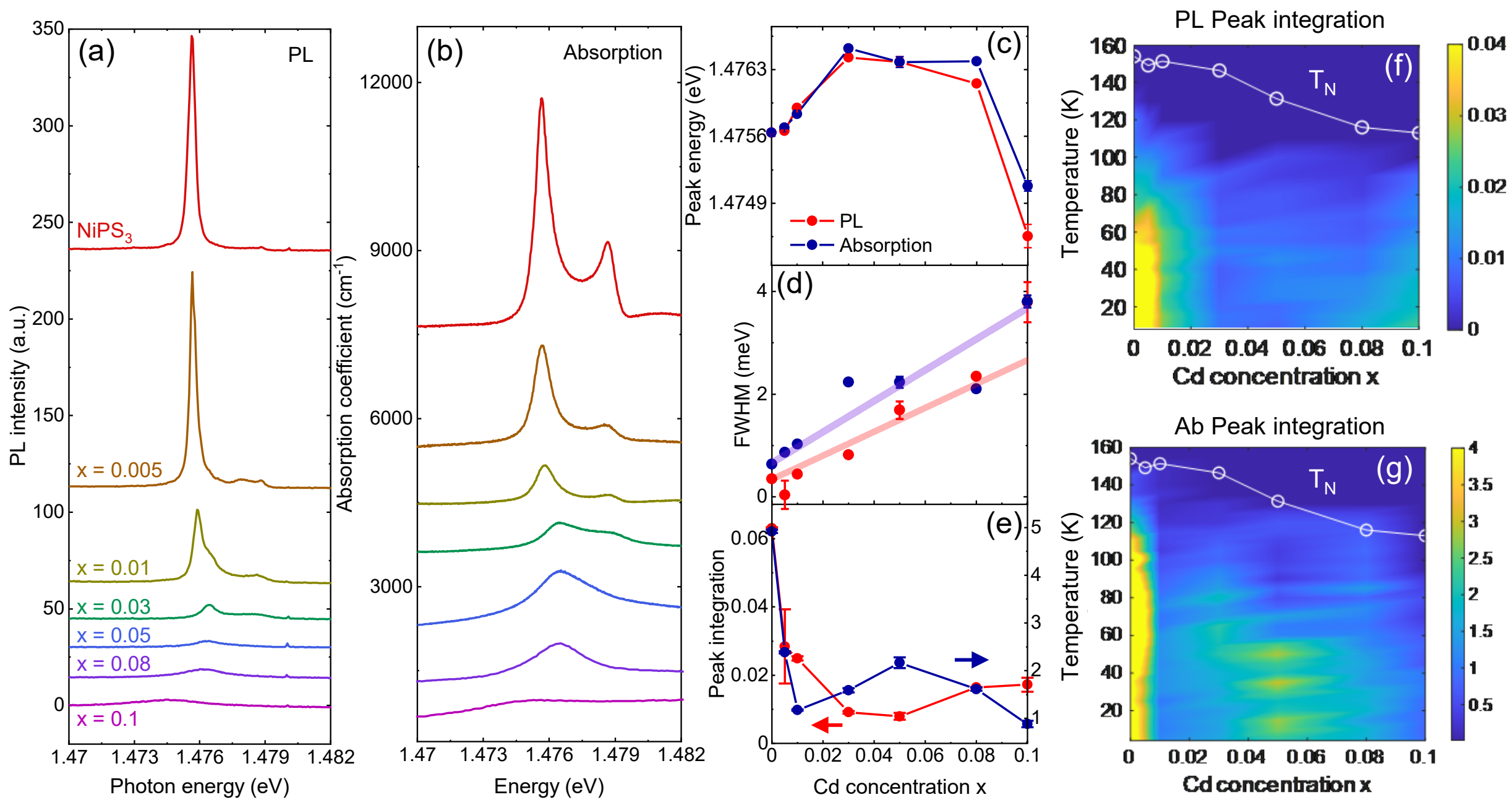

**Figure 3.** (a) PL spectra as a function of Cd concentration x at 8 K. (b) Absorption spectra as a function of Cd concentration x at 4 K. (c) Peak energy, (d) FWHM, and (e) integrated peak of the fitted main exciton peak of PL and absorption. (f-g) The color plot of the temperature-Cd concentration mapping of the integrated peak for the doped system by PL and absorption spectra. The white scatter and the line indicate the Neel temperature obtained by the magnetic susceptibility.

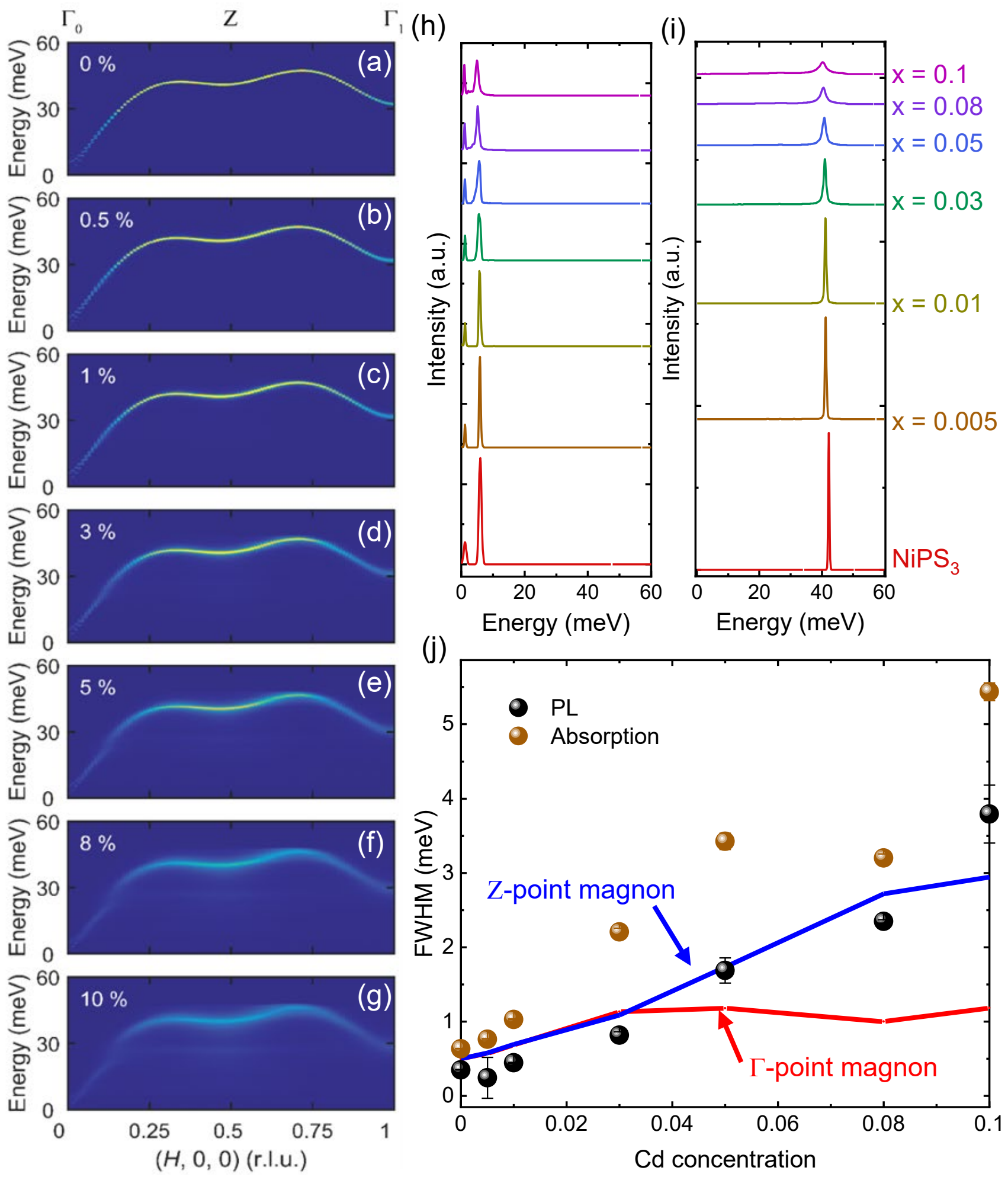

**Figure 4.** (a-g) Theoretical simulation of the magnon dispersion for $Ni_{1-x}Cd_xPS_3$ systems along *(H,0,0)* direction. (h-i) Q-cuts at $\Gamma_0 = (0,0,0)$ and Z = (0.5, 0, 0) of the doped systems (j) FWHM of magnon modes from the theoretical calculation, where the red line is a Γ–point magnon, and the blue line is a Z–point magnon and FWHM of PL and absorption at the base temperature as a function of a Cd concentration.

## Table of Contents

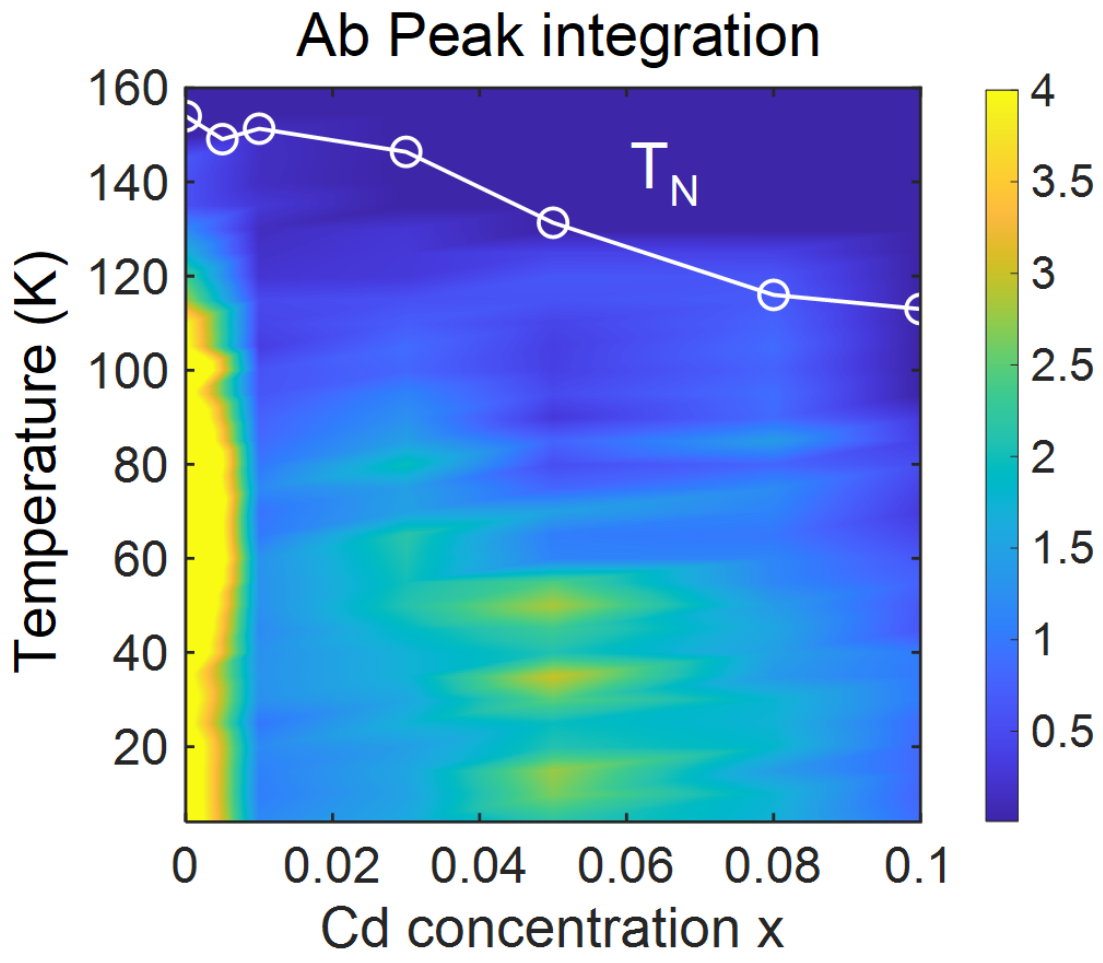

# Supporting Information for

# Rapid suppression of quantum many-body magnetic exciton in doped van der Waals antiferromagnet (Ni,Cd)$PS_3$

Junghyun Kim[1,2#], Woongki Na[3#], Jonghyeon Kim[4#], Pyeongjae Park[1,2], Kaixuan Zhang[1,2], Inho Hwang[1,2], Young-Woo Son[5], Jae Hoon Kim[4,*], Hyeonsik Cheong[3,*], and Je-Geun Park[1,2,*]

[1]Center for Quantum Materials, Seoul National University, Seoul 08826, Republic of Korea

[2]Department of Physics & Astronomy, Seoul National University, Seoul 08826, Republic of Korea

[3]Department of Physics, Sogang University, Seoul 04107, Republic of Korea

[4]Department of Physics, Yonsei University, Seoul 03722, Republic of Korea

[5]School of Computational Sciences, Korea Institute for Advanced Study, Seoul 02455, Republic of Korea

# Authors with equal contribution

* Corresponding author: super@yonsei.ac.kr; hcheong@sogang.ac.kr; jgpark10@snu.ac.kr

**Section S1. Sample preparation and characterization**

The doped crystals $Ni_{1-x}Cd_xPS_3$ were synthesized by the solid-state reaction (SSR) for 5 days and the following chemical vapor transport (CVT) method for 7 days. The sealed quartz evacuated by $4\times10^{-2}$ torr with starting materials Ni (99.999%), Cd (99.999%), P (99.999%) and S (99.999%) powders in the stoichiometric ratio was placed in the 2-zone furnace (Ajeon Instrument, Korea). The synthesis temperature for SSR is set at 700°C for x = 0, 0.005, 0.01, 0.03, and 0.05, while we set it at 740°C and 770°C for x = 0.08 and 0.1, respectively. The 2-zone furnace temperature for CVT is set at 720°C / 750°C in the cold/hot zone for x = 0, 0.005, 0.01, 0.03, and 0.05, while it was set at 760°C /790°C and 790°C /820°C for x = 0.08 and 0.1, respectively. The obtained crystals are characterized by SEM-EDX (EM-30, COXEM) and powder XRD (Rigaku). The magnetic susceptibility was measured by the Magnetic Properties Measurement System (MPMS-XL EverCool, Quantum Design). The micro-Raman measurement for 2D mapping was performed by the Confocal Raman spectroscopy (XperRam, NanoBase, 532 nm laser) at room temperature. The nanoflake was exfoliated by the mechanical method using Scotch tape.

## Section S2. Optical Spectroscopy – Optical absorption measurement

The optical transmittance measurements were conducted on a grating spectrophotometer (Cary 5000, Agilent) over the energy range of 0.62 to 1.67 eV at a resolution of about 0.155 $cm^{-1}$ with a spectral bandwidth of about 0.3 nm. The temperature varied over 4 to 290 K using a He-free optical cryostat (Cryostation, Montana Instruments). The samples were mechanically exfoliated down to tens of micrometres, and they were mounted on Au-coated Cu sample holders with a diameter of 1 mm and attached by Kapton tape. The microscopic absorption measurements were also conducted on a microscope spectrophotometer (2030PV, Craic) over the energy range 0.59 to 2.48 eV at a resolution of 2 nm. The few-layer samples were prepared mechanically and mounted on a sapphire substrate.

**Section S3. Optical Spectroscopy – PL and Raman measurement**

Temperature-dependent photoluminescence (PL) spectra of bulk $NiPS_3$ and Cd-doped $NiPS_3$ were measured using a closed-cycle He cryostat in a macro-measurement system. The 632.8 nm (1.96 eV) line of a He-Ne laser was used as the excitation source for all the measurements. The excitation laser was focused by a spherical lens (f=75 mm) to a spot of ~50 μm with a power of 1 mW. The emitted signals were dispersed by a Jobin-Yvon Horiba iHR550 spectrometer (1,200 grooves per mm) and detected with a liquid-nitrogen-cooled back-illuminated charge-coupled-device (CCD) detector. Raman measurements used an analyzer and a half-wave plate to discriminate the scattered light. The analyzer was set such that photons with polarization parallel or cross to the incident polarization pass through. Finally, an achromatic half-wave plate was placed in front of the spectrometer to keep the polarization direction of the signal entering the spectrometer constant with respect to the groove direction of the grating.

## Section S4. Theoretical calculations

To calculate the spin configuration and spin dynamics of $Ni_{1-x}Cd_xPS_3$, we introduced randomly distributed vacancies into a 20 × 20 × 1 supercell of $NiPS_3$ (= 1600 Ni sites) with periodic boundary conditions. To simulate this spin system, we adopted the spin Hamiltonian of pristine $NiPS_3$ determined by high-quality single-crystal inelastic neutron scattering data[1]. For simplicity, we neglected inter-layer couplings. First, a magnetic ground state of the diluted system was examined by the energy minimization algorithm, which resulted in no change in the magnetic structure by vacancies. After that, we diagonalized the spin Hamiltonian of the 20 × 20 × 1 spin system using linear spin wave theory and calculated its dynamical structure factor using the SpinW library.[2] To properly consider Cd impurities' randomness, we calculated the dynamical structure factor over 30 independent spin systems with different vacancy distributions and used the averaged structure factor. The calculated spectra were convoluted by the Gaussian function along the energy axis, with its FWHM being 0.5 meV. This is similar to the FWHM of the measured exciton peak in pristine $NiPS_3$, which is considered the instrumental resolution of our optical system. Thus, this convolution enables a fairer comparison between the FWHMs of calculated magnon spectra and our measured optical spectroscopy data. This broadening is included in all plots of Fig. 4.

**Section S5. Raman spectroscopy of doped systems**

The macro-Raman at the base temperature of 8 K was measured for the bulk crystal of all the doped systems (Figure S1). The original Raman peak of the pure $NiPS_3$ is maintained in the doped systems, showing a gradual redshift of the peak position. Depending on the polarisation direction, the small splitting of the P2 peak, where this specific peak is a satellite of magnetism, is also maintained in the doped systems. We can also detect the clear PL peak (Figure 3) in the same position where the splitting of the P2 peak is apparent. We observe all the peak positions of P2. P3 and P4 shift as the same amount in each system. The increasing tendency of small backgrounds, except for the main peaks in the overall spectra at a high doping regime of parallel polarization, is needed for further study.

To check the homogeneity of the pure and doped crystals, we also measured the micro-Raman spectra of doped systems inside the glove box ($O_2$ < 0.1 ppm, $H_2O$ < 0.1 ppm) at room temperature (Figure S2). The nanoflake is exfoliated on the $SiO_2$ substrate by Scotch tape in the air and moved inside the glove box. The 2D mapping of Raman spectra is performed by the program (NanoSpectrum, Xper-PC, NANOBASE) with a 2D step size of 1 μm × 1 μm. The 2D mapping image is drawn for every system's peak, P2. We see that the intensity of color mapping of doped systems is relatively homogeneous as that of a pure system within the same thickness region. Finally, we note that the intensity of peak intensity changes in the different thickness regions; thus, the color of intensity is different in the different thickness regions.

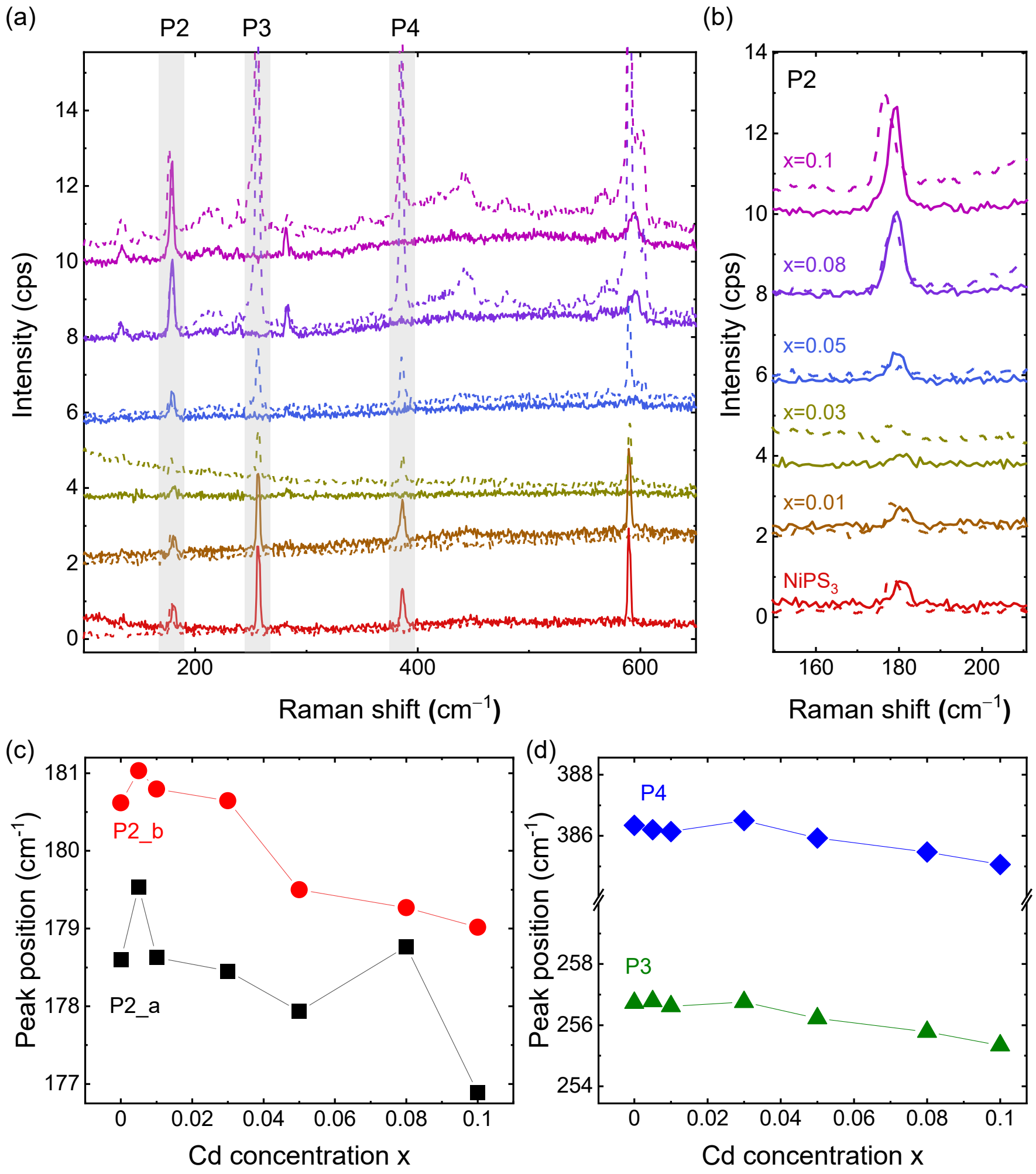

**Figure S1.** (a) Macro-Raman spectra of the doped systems measured at T = 8 K in parallel (solid line) and cross (dashed line) polarization scattering configuration. (b) Dopant ratio dependence of the magnetic-order-induced frequency P2 as a function of Cd dopant concentration (c) Peak positions of P2 where P2_a indicates the lower frequency of polarization direction and P2_b indicates the higher frequency of polarization direction (d) Peak positions of P3 and P4 as a function of Cd dopant concentration.

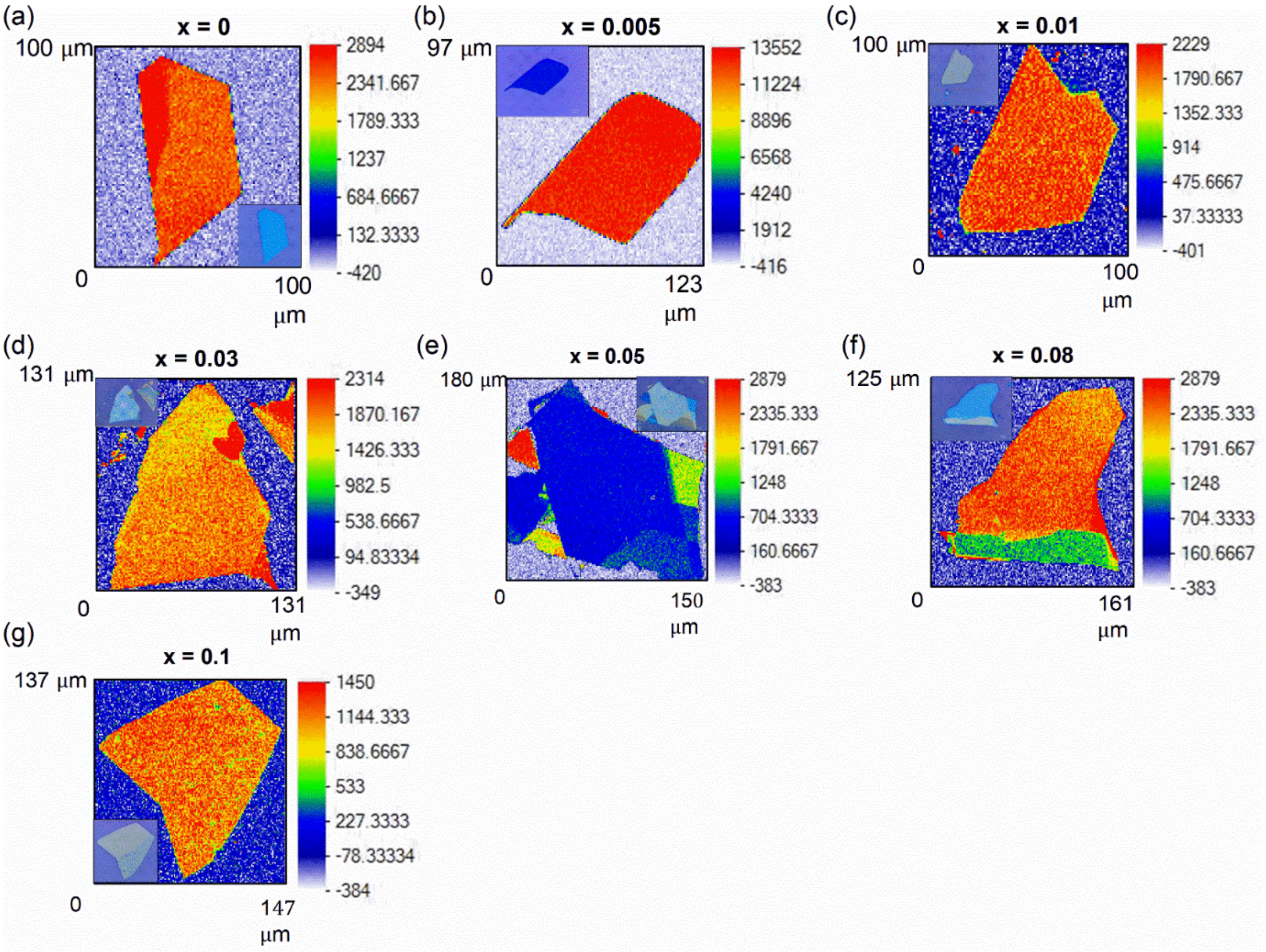

**Figure S2.** (a)-(g) Micro-Raman mapping image from x = 0 to x = 0.1. The mapping step along the x and y axis is 1 μm. The Inset image indicates the optical image observed by the microscope.

## Section S6. Derivatives of the magnetic susceptibility

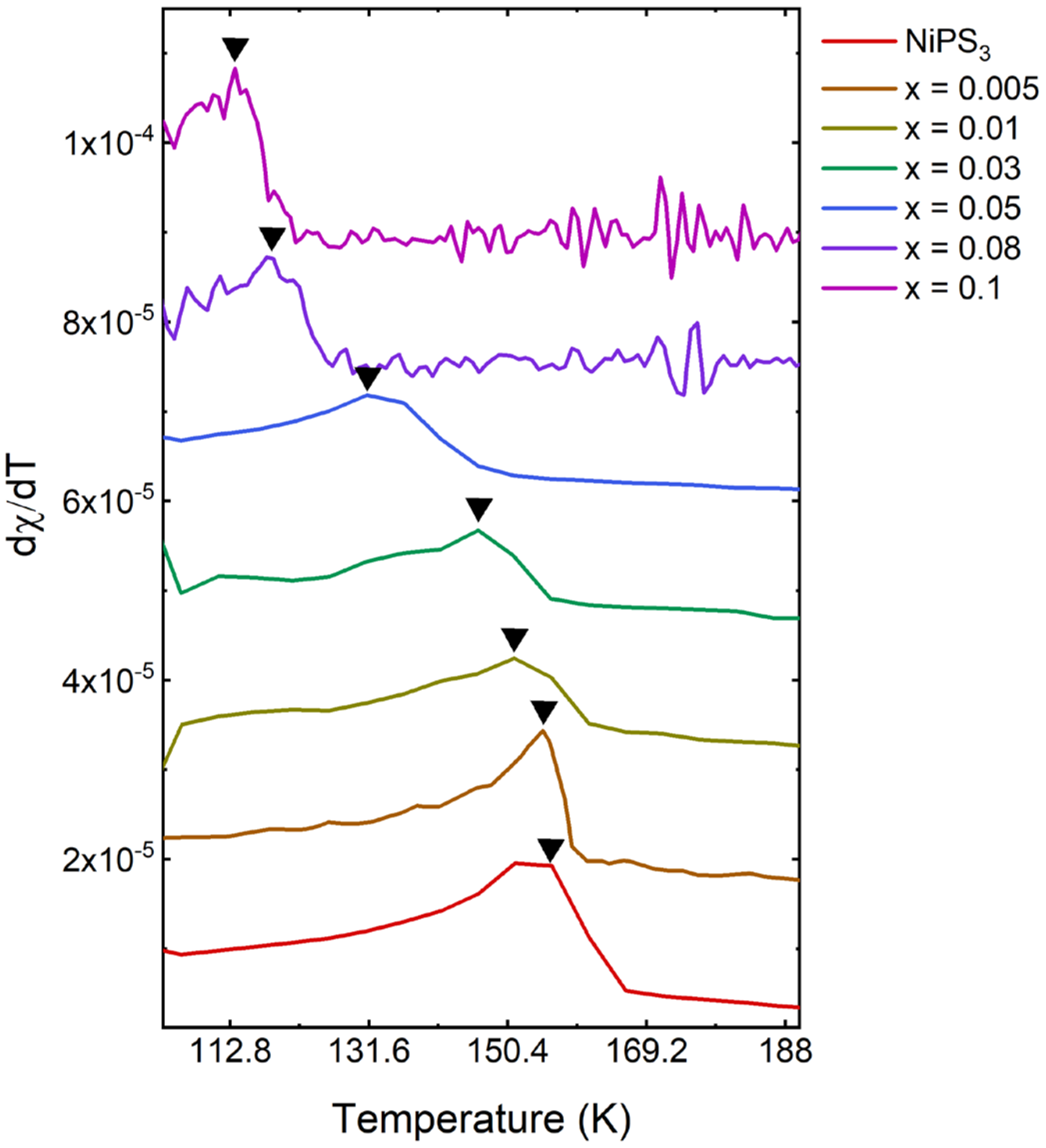

**Figure S3.** The derivatives of the magnetic susceptibility are shown in Figure 2. The peak in the derivatives is indicated by a black marker, which is the position of the Neel temperature of each system.

## Section S7. Temperature dependence of PL and absorption spectra of doped systems

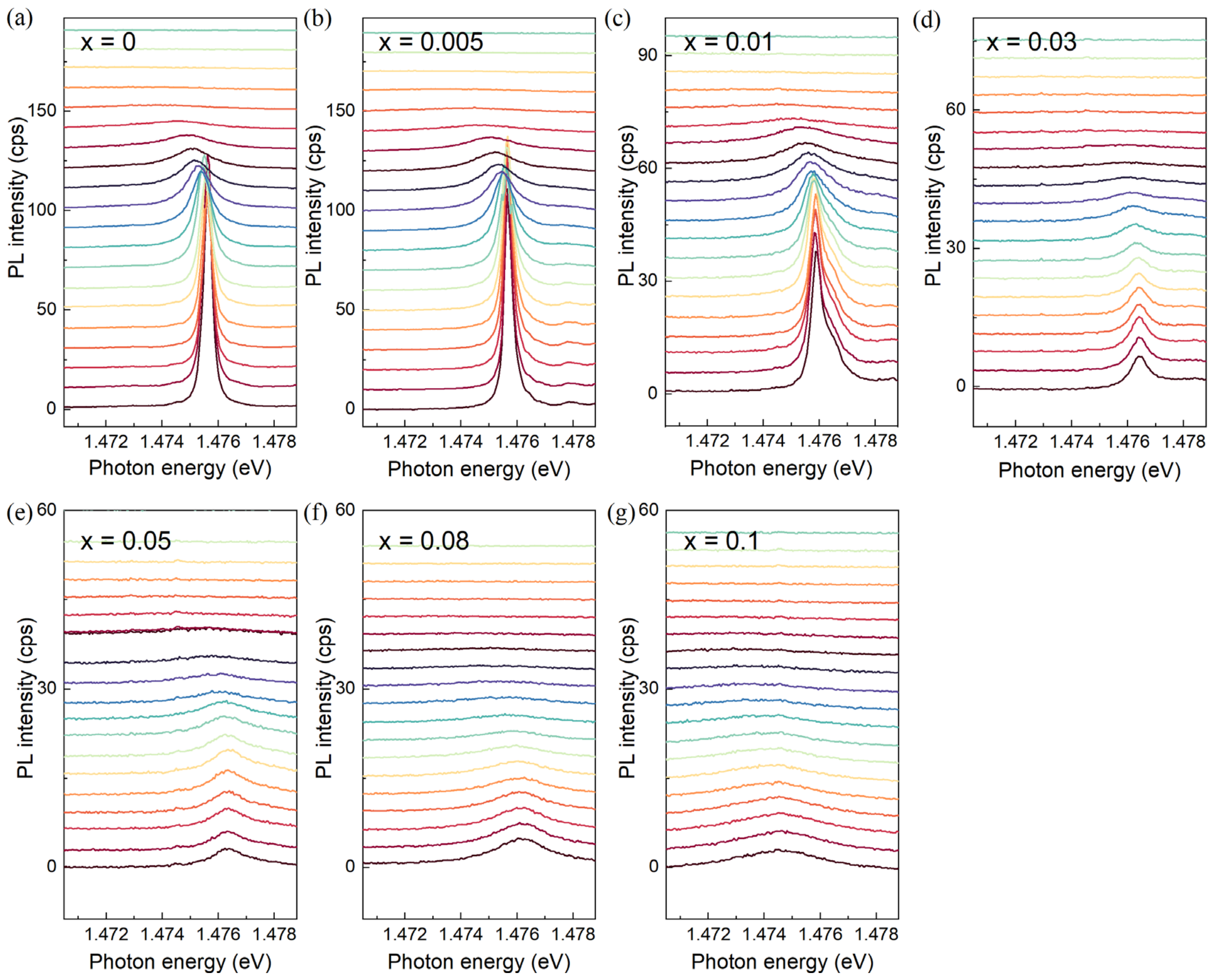

**Figure S4.** (a)-(g) Temperature dependence of PL spectra from 8 K to 130 K for $Ni_{1-x}Cd_xPS_3$ systems. The plot is stacked, and the temperature decreases from top to bottom.

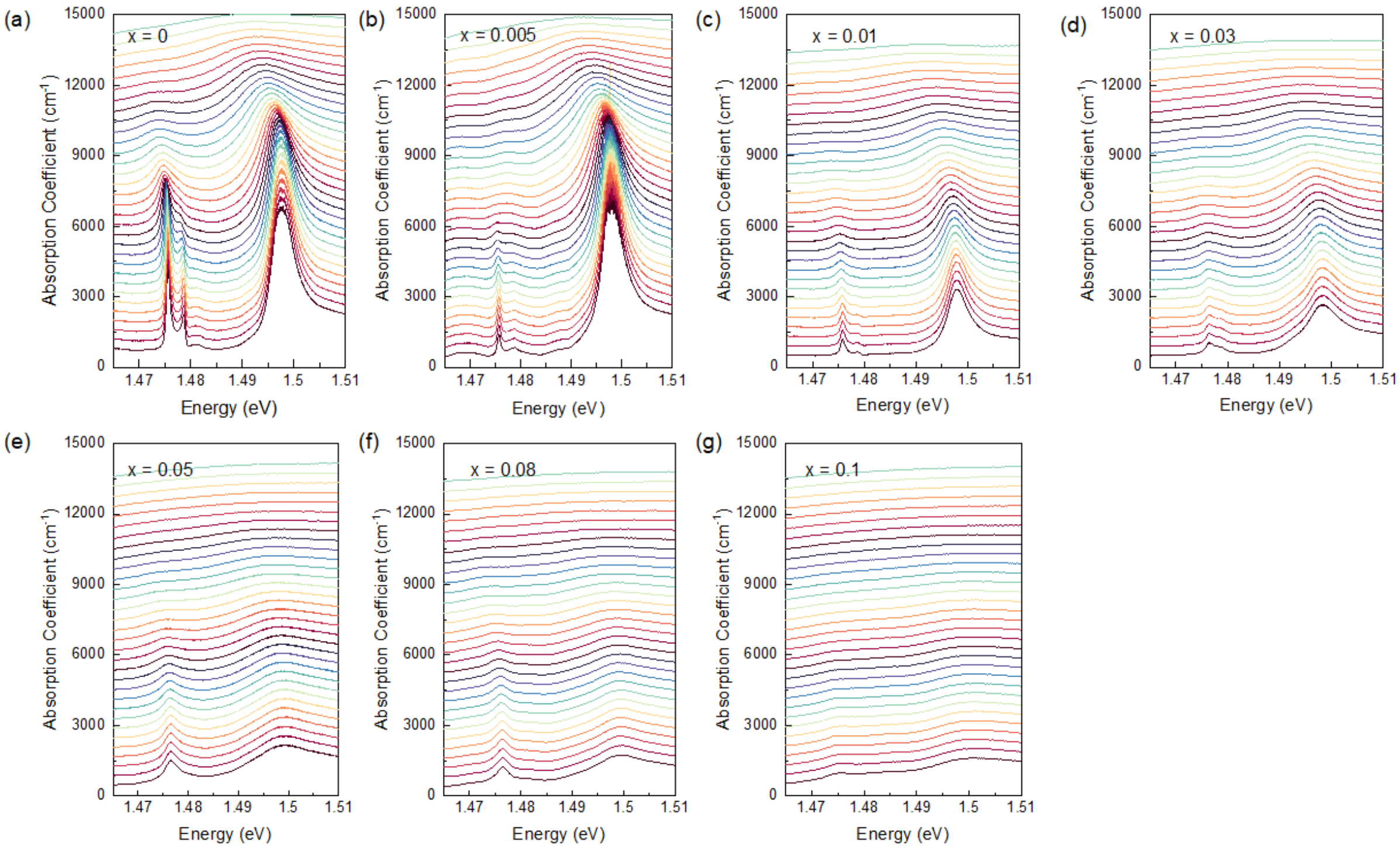

**Figure S5.** (a)-(g) Temperature dependence of absorption spectra from 4 to 160 K for $Ni_{1-x}Cd_xPS_3$ systems. The plot is stacked, and the temperature decreases from top to bottom.

**Section S8. Fitted exciton peak of PL and absorption spectra of doped systems**

The fitting data of PL and absorption spectra at the base temperature are shown for all the systems in Figures S6 and S7, respectively. The Lorentzian function fits the central exciton pea, and the Gaussian function fits other sidebands. We extract the parameters of the central exciton peak with the largest amplitude. We used a few more sidebands for the best fit. As a dopant concentration increases, the central exciton peak merges with sidebands, i.e. the peak positions get closer. Thus, those peaks become a single peak at a high doping regime, which requires further investigation.

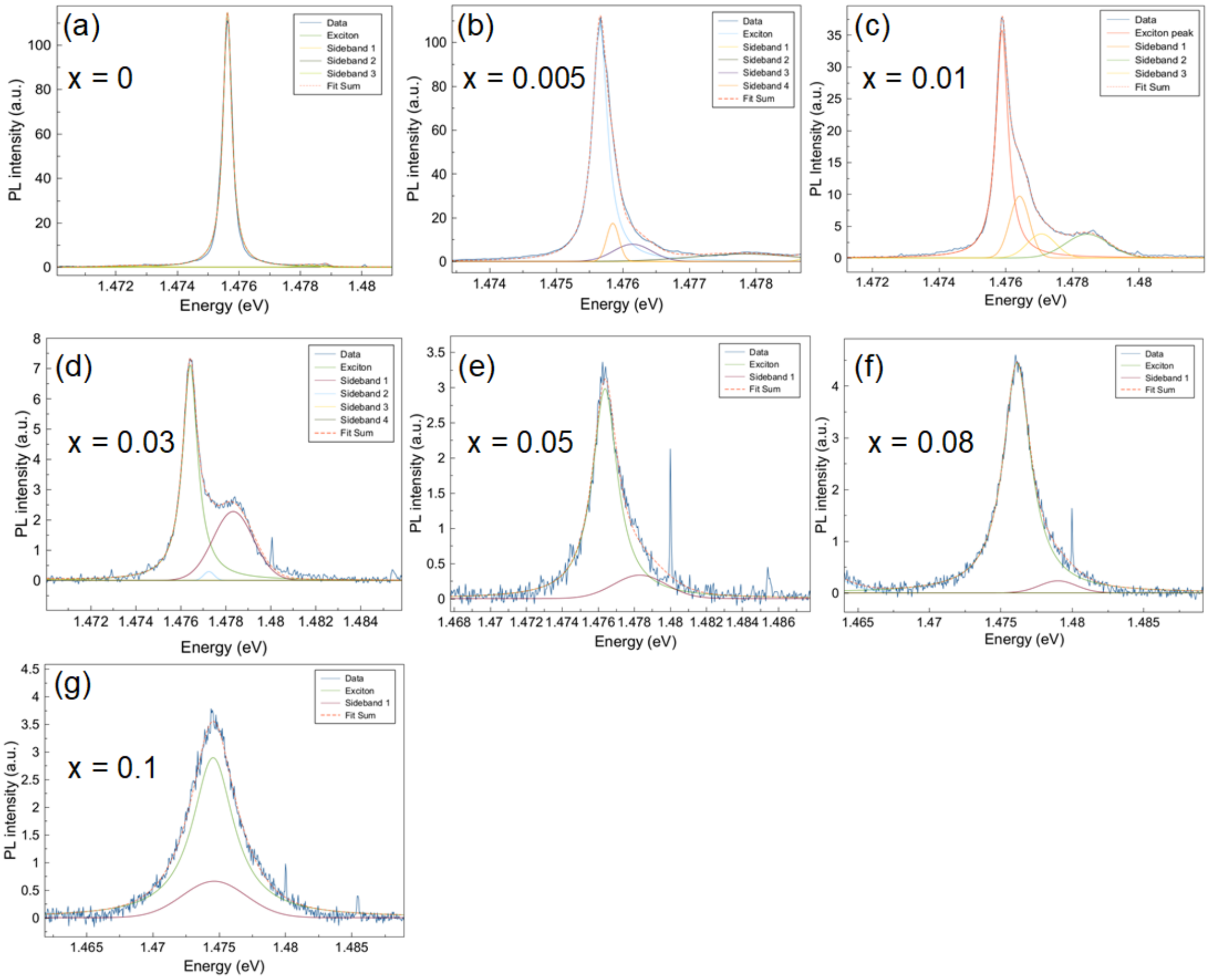

**Figure S6.** (a)-(g) Fitted exciton peak and sidebands of PL spectra for all doped systems. The Lorentzian function fits the central exciton peak.

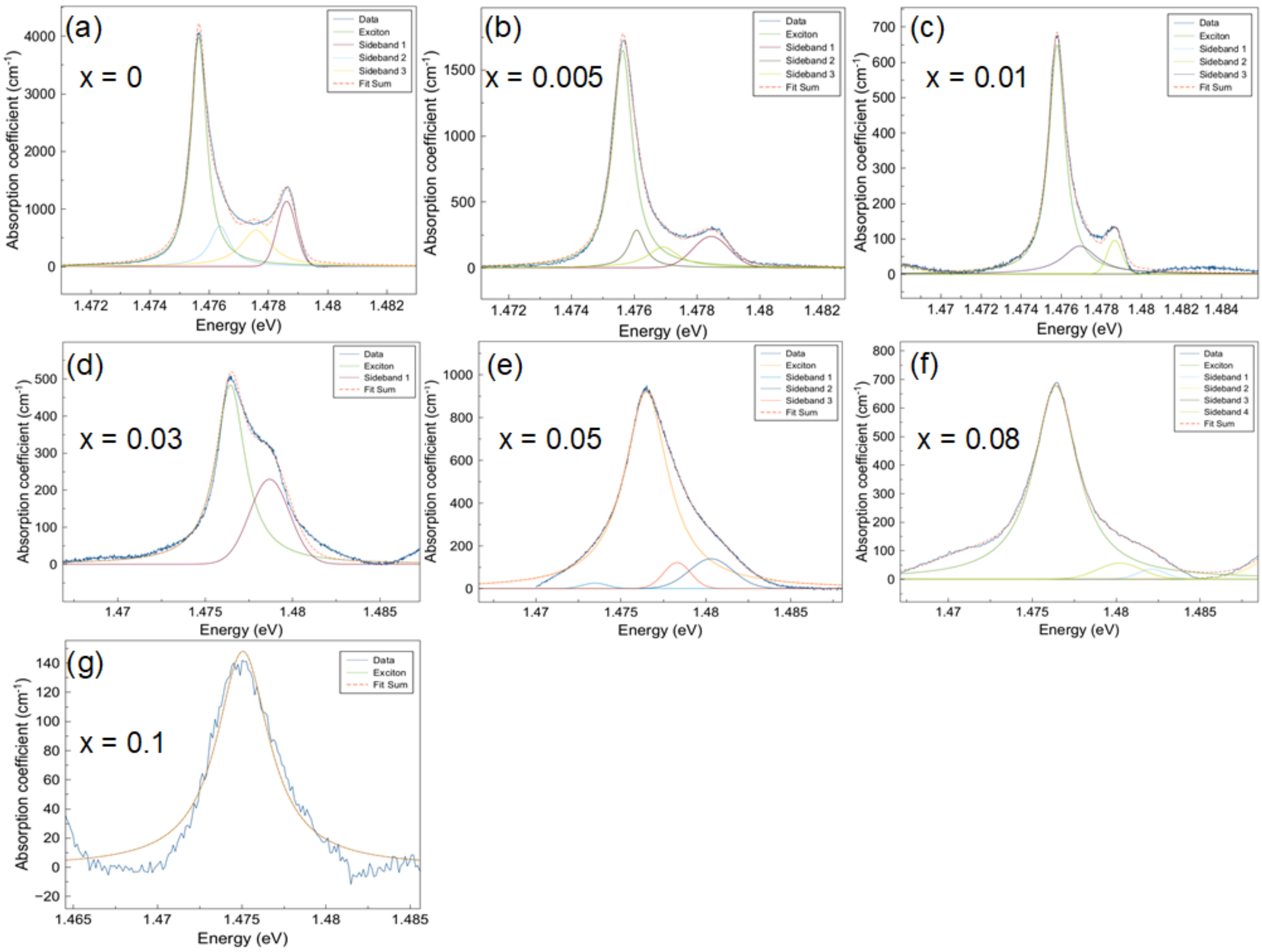

**Figure S7.** (a)-(g) Fitted exciton peak and sidebands of absorption spectra for all doped systems. The Lorentzian function fits the central exciton peak.

## Section S9. Laser power dependence of PL

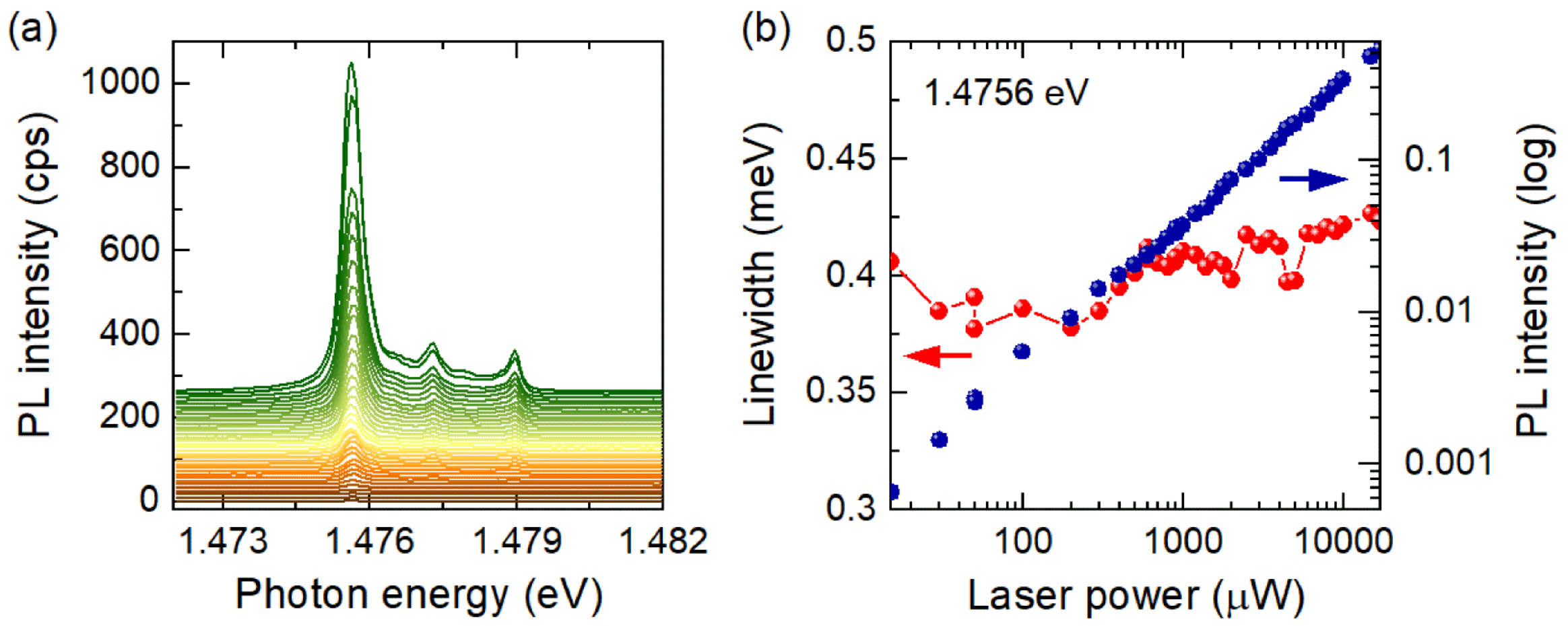

**Figure S8.** (a) Full spectra of laser power dependence of PL intensity. (b) Linewidth and the PL intensity as a function of laser power from 15 μW to 17200 μW.

## Section S10. Comparison of Neel temperatures

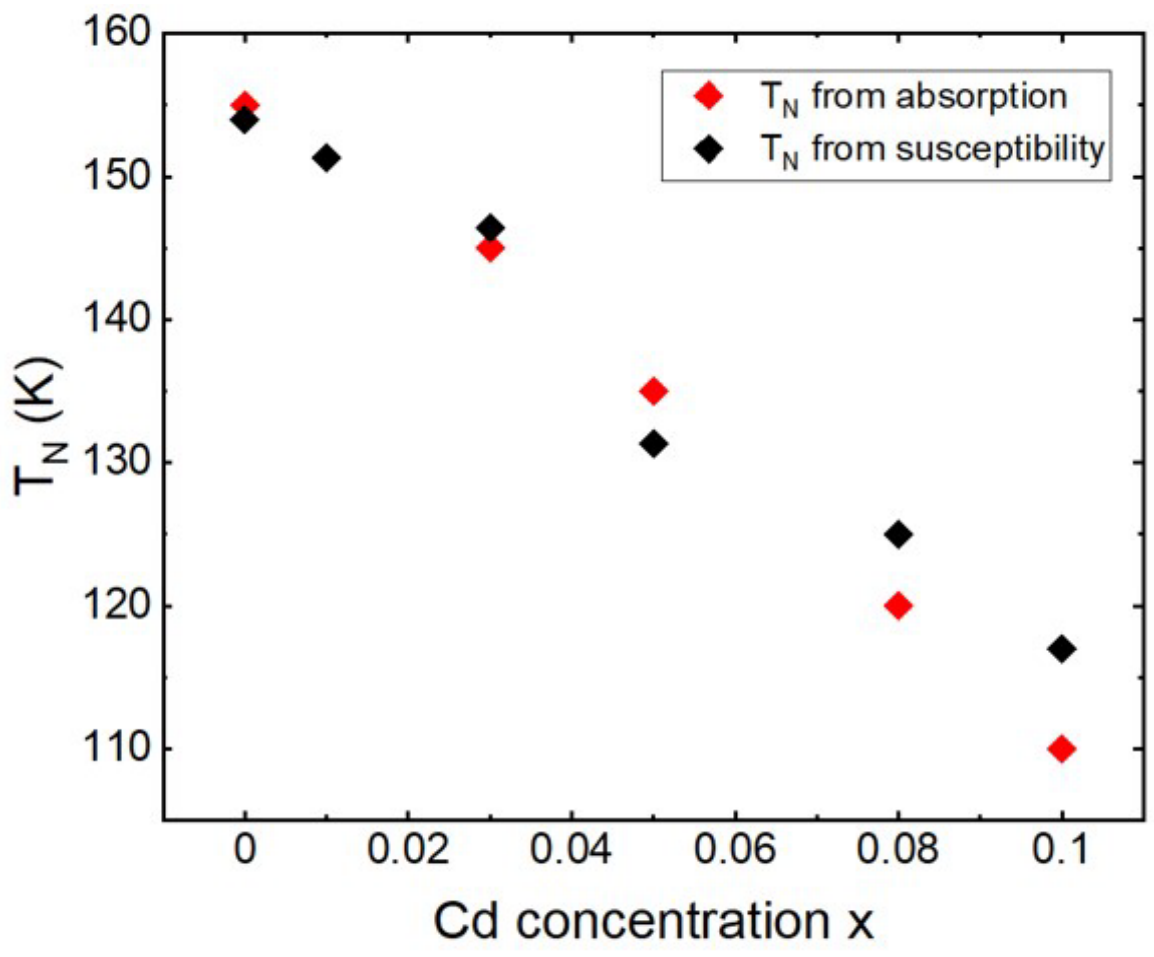

**Figure S9.** Comparison for the Neel temperatures, $T_N$, obtained from the magnetic susceptibility and the optical absorption spectra. The value of $T_N$ obtained from the optical absorption means the temperature where the magnetic exciton peak disappeared.

**Section S11. A color plot of FWHM as a function of temperature and dopant concentration**

We extract the FWHM data of the exciton peak by fitting the PL and absorption spectra for all the temperatures. The color plots of FWHM are drawn as a function of temperature and the dopant concentration for PL and absorption. The Neel temperature data, obtained from the magnetic susceptibility, is plotted to compare the exciton suppression and the Neel temperature decrement. The bright region indicates that the exciton peak has a much narrower linewidth than a darker region, another sign of coherence. Contrary to the color plot of amplitude (Figure 3f-g), the bright region of FWHM fades more slowly than that in the case of amplitude. The suppression speed of the bright region of FWHM is similar to that of Neel temperature as a dopant concentration increases.

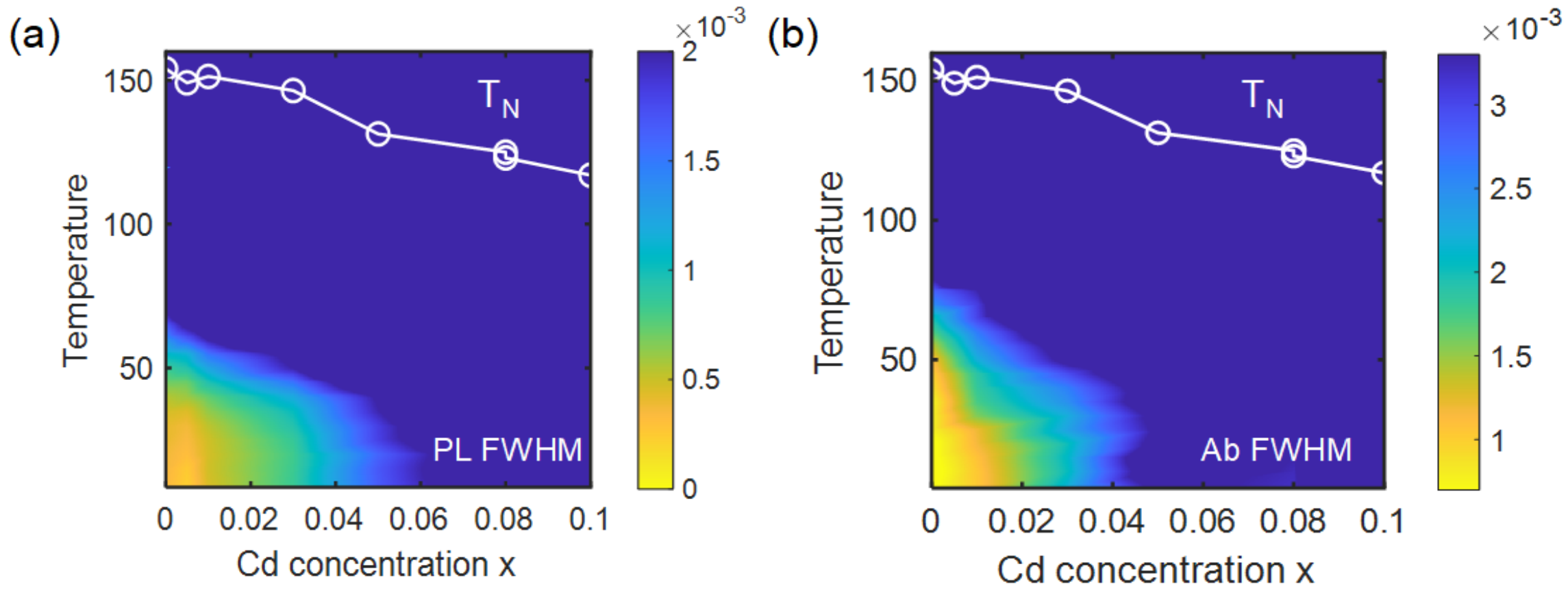

**Figure S10.** Color plot the FWHM for the doped system by PL and absorption spectra as a function of temperature and the dopant concentration. White scatter and line indicate the Neel temperature obtained from the magnetic susceptibility data (Figure 2b).

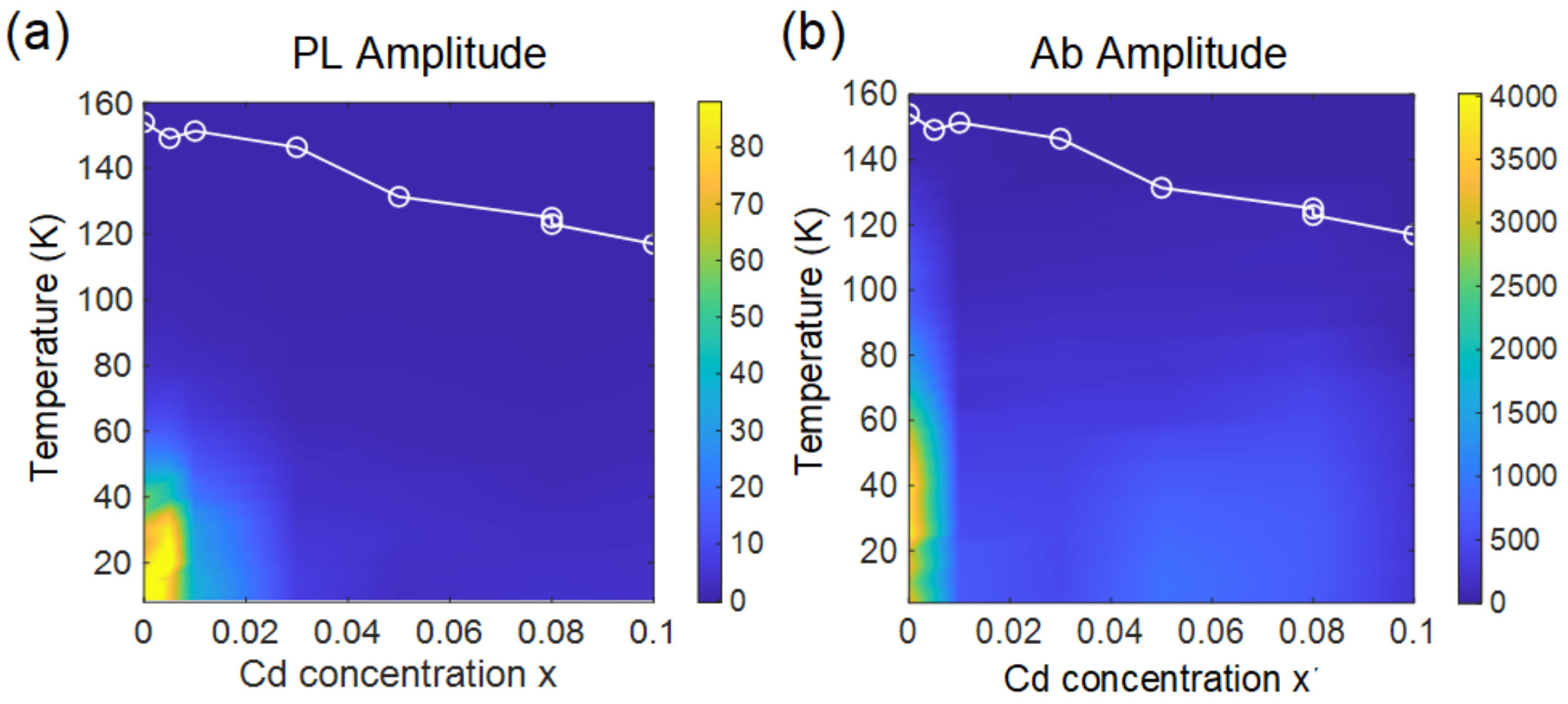

**Figure S11.** Color plot the amplitude for the doped system by PL and absorption spectra as a function of temperature and the dopant concentration. White scatter and line indicate the Neel temperature obtained from the magnetic susceptibility data (Figure 2b).

**Section S12. Simulation of the ground-state spin configuration of doped systems**

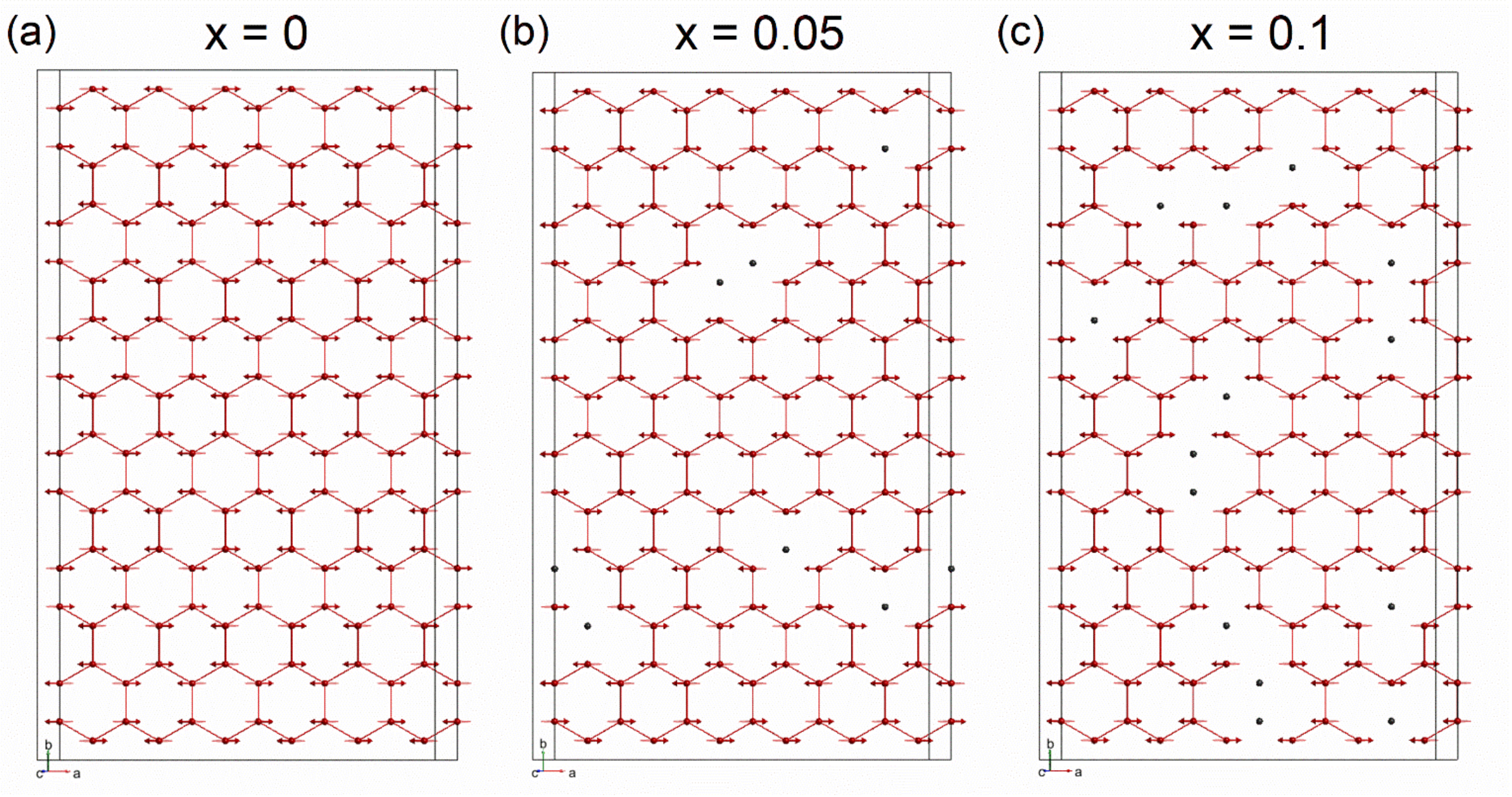

**Figure S12.** Simulated spin configuration for each ground state of $Ni_{1-x}Cd_xPS_3$ systems by spinW program for (a) x = 0, (b) x = 0.05, and (c) x = 0.1. The red arrow indicates Ni spins on the honeycomb lattice, and dark blue dots indicate the Cd atoms in the Ni spin-lattice. The magnon band dispersions are also calculated for x = 0.005, 0.01, 0.03, and 0.08.

**S13. Schematics of magnetic exciton in the doped system**

The schematics in Fig. S12 of the magnetic excitons show the original bright exciton, distributed bright exciton, and optically inactive dark exciton. The Frenkel exciton in $NiS_6$ is a tightly bound exciton on one lattice site. The schematics show that the original bright exciton is localized at the Ni site. The large coherent amplitude and narrow linewidth have been produced from the enormous exciton energy, corresponding to the energy difference between the Zhang-Rice states. Thus, the original exciton energy can be assumed as a delta function, $U_{ex, original}$. The Cd site has no exciton from the charge-transfer-like Ni cluster; thus, the original Zhang-Rice states become absent. Therefore, the $CdS_6$ cluster and the intralayer nearest $NiS_6$ cluster, which directly bonds with the $CdS_6$ cluster, do not have a bright exciton as the original bright exciton. Additionally, the interlayer exciton on $NiS_6$ without direct bonding with the $CdS_6$ cluster can still be bright. However, it is possible to have a distributed energy potential due to Cd atoms, where the distance to the dopant is enough to affect the exciton state. Therefore, the exciton energy that arises from this bright exciton will also have a distribution so that the linewidth of the magnetic exciton peak is broadened.

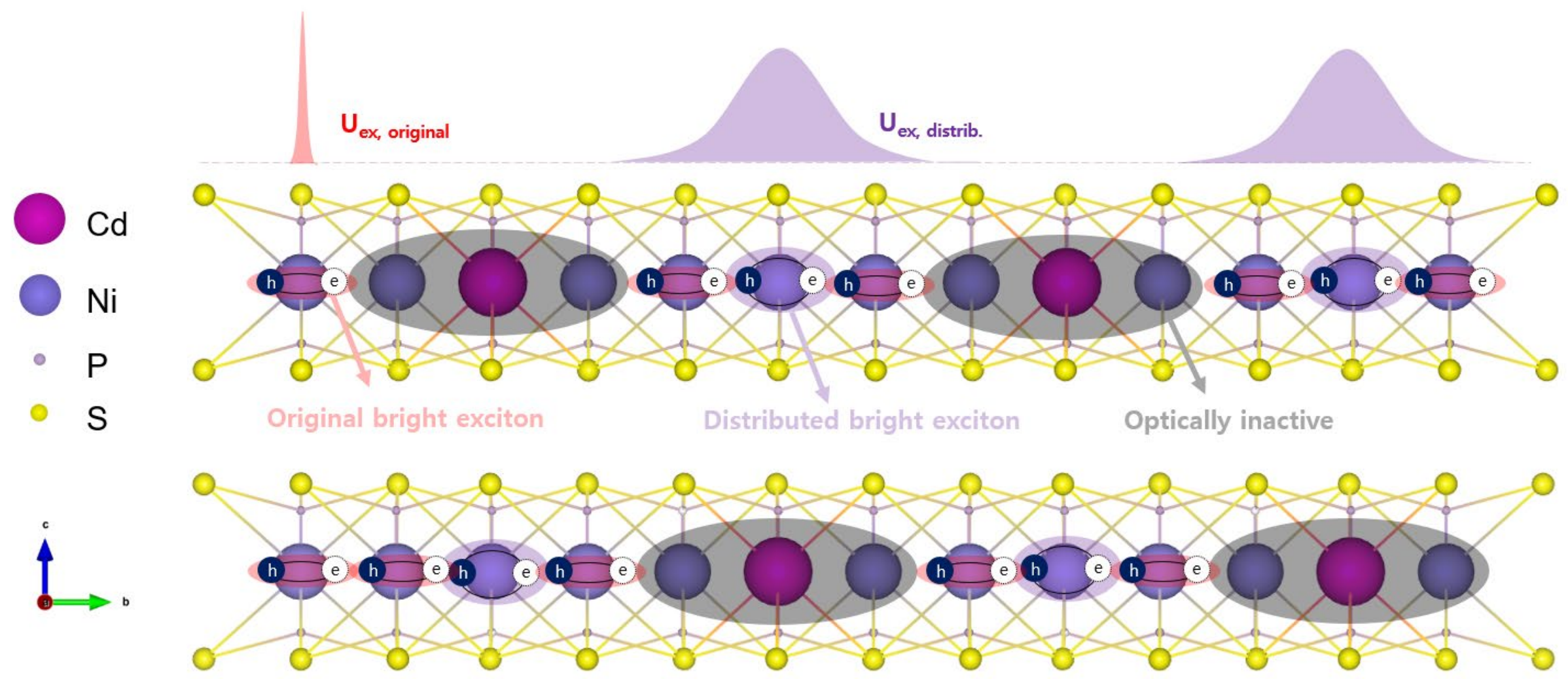

**Figure S13. Schematics of the types of magnetic exciton in the doped system.**